\documentstyle[a4,epsfig,12pt]{article}
\begin{document}
 
%%%%%%%%%%%%%%%%%%%%%
\newdimen\picraise
\newcommand\picbox[1]
{
  \setbox0=\hbox{\input{#1}}
  \picraise=-0.5\ht0
  \advance\picraise by 0.5\dp0
  \advance\picraise by 3pt      % correct for height of `=' above baseline
  \hbox{\raise\picraise \box0}
}
%%%%%%%%%%%%%%%%%%%%%

\title{ORSAY LECTURES ON CONFINEMENT(III)}
\author{V. N. Gribov\\
     L.D. Landau Institute for Theoretical Physics, Moscow \\
     and\\
     KFKI Research Institute for Particle and Nuclear Physics, Budapest \\
     and\\
     Laboratoire de Physique Th\'eorique et Haute Energies \\
     Universit\'e de Paris XI, b\^atiment 211, 91405 Orsay Cedex, France}
\date{April 23, 1992}
\maketitle
\vspace*{-120mm}{\hfill {\sc LPT-Orsay-99-37}\par \hfill hep-ph/9905285}

\newpage
\section*{Light quark confinement\footnote{This is the third lecture on quark 
confinement given by V.N.Gribov in 1992 in Orsay. An extensive discussion of the
consequences of all this for the structure of the Green function can be found in
[5,6] - in the two last papers concluding his 20 years long study of the problem
of quark confinement in QCD.}
\footnote{The text was prepared for
publication by Yu. Dokshitzer, B. Metsch and J. Nyiri on the basis of a tape    
recording and notes taken during the lecture}}
We have described the confinement of heavy quarks in an analogy with 
the theory of the supercharged nucleus [1,2]. Let us now suppose again
that $\alpha$ is behaving like
\[ \input{O1.pstex_t}  \]
Making this assumption, we are neglecting gluon-gluon interactions and the
existence of gluons as real particles. Our aim is to see, what can arise
from the discussion of light quarks only. We introduce $\lambda$
corresponding to $\alpha_c$ and consider quark masses $m_0 \ll \lambda$.
The interactions of light quarks (for which $m_0 \ll \lambda$) will be
discussed in a rather simplified way. We will take into account all
possible interactions 
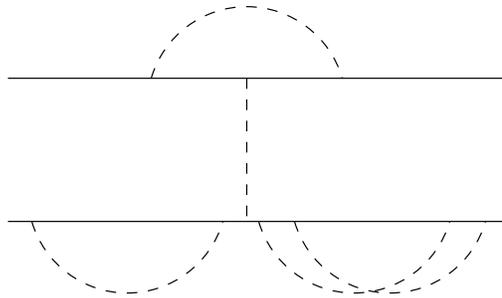
\[ \begin{picture}(0,0)%
\epsfig{file=O2.pstex}%
\end{picture}%
\setlength{\unitlength}{0.00083300in}%
\begingroup\makeatletter\ifx\SetFigFont\undefined%
\gdef\SetFigFont#1#2#3#4#5{%
  \reset@font\fontsize{#1}{#2pt}%
  \fontfamily{#3}\fontseries{#4}\fontshape{#5}%
  \selectfont}%
\fi\endgroup%
\begin{picture}(3174,1818)(1189,-2620)
\end{picture}
 \]
where the gluon propagator (considered as an effective photon),
corresponding to the dotted line is 
%---   where $D_{\mu\nu}$ corresponding to the effective photon line is, by
%---   definition, $\frac{\alpha}{q^2}\delta_{\mu\nu}$:
\begin{equation}
\label{1}
D_{\mu\nu} = \frac{\alpha}{q^2}\delta_{\mu\nu} .
\end{equation}
Further, we look for a model which enables us to see, what happens to
the fermions if there is an interaction 
%---   $\alpha$  as supposed.
between them, as indicated above. The
question is, how the bound states or the Green function behave in such
a case.

Let us consider the energy of two quarks, $u$ and $\overline{d}$, for
example. Without interaction there will be positive energy states with
$E > 2m$
and negative energy states with $E<-2m$:
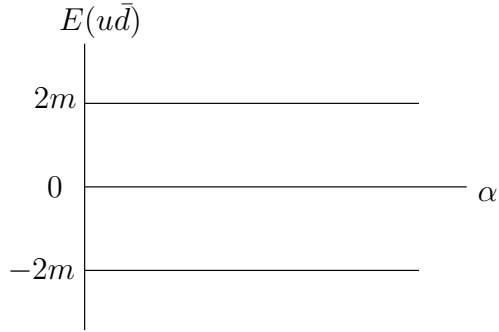
\[  \input{O3.pstex_t} \]
Introducing the interaction, for small $\alpha$ we will
find that there are some bound states near $2m$ and $-2m$.
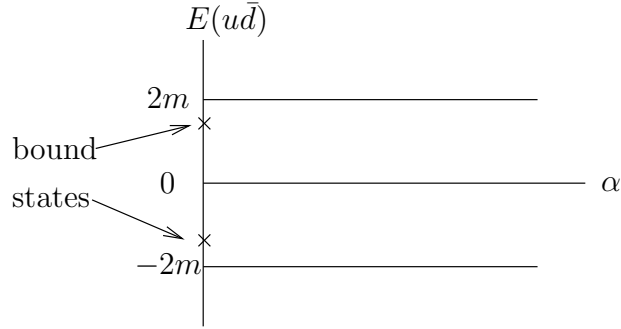
\[  \input{O4.pstex_t}  \]
So far, we consider the usual Dirac vacuum: the negative 
energy
states are
occupied, and the positive ones are empty. Increasing the coupling, i.e.
increasing $\alpha$, we could expect that the 
%---   binding 
magnitude of the energy is decreasing
and the levels corresponding to the bound states will come closer and
closer to zero.
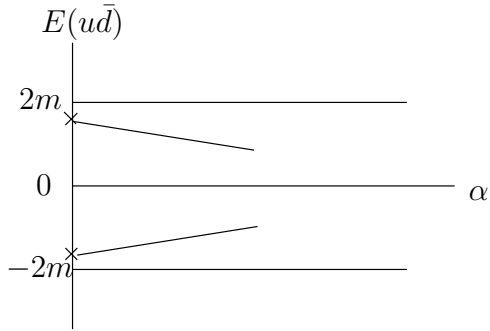
\[  \input{O5.pstex_t}  \]
With a further increase of the coupling up to a critical value, one
possibility for the levels will be just to approach the zero line and
never cross. There is, however, also a possibility of crossing.
%---   to cross this line. 
We will see that the first case corresponds to normal spontaneous
symmetry breaking. But, if the levels cross, and especially in the most
clear case, when they pass the lines $2m$ and $-2m$, respectively,
everything will change and we arrive at
%---   come to 
very different phenomena:
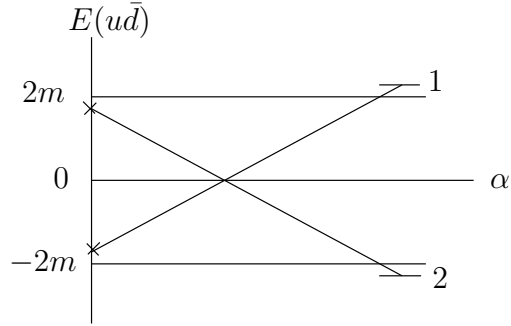
\[ \input{O6.pstex_t}  \]
Indeed, now the 
original 
vacuum which corresponds to the case when level $2$ is
empty and level $1$ is occupied, is absolutely unstable. We have to
fill the new negative energy state and leave the positive energy
level empty. But by filling this new state, we get an excitation, a meson-type
state with a mass $\mu$. For free quarks this would mean that the quark
with negative energy decays into a negative energy meson (filling the
negative energy levels) and creates a positive energy quark. As a result,
the Dirac picture in which all negative energy levels are filled up and
all positive energy levels are empty, is destroyed. But if so, a positive
energy quark also decays into a positive energy meson and a quark with
negative energy. This means that both decays
\begin{eqnarray*}
  q^- & \rightarrow & \mu_- + q^+ \\
  q^+ & \rightarrow & \mu_+ + q^-
\end{eqnarray*}
are possible, and both $q^-$ and $q^+$ are unstable.

The question is now, how to deal with the bound state
%---   this 
problem. Of course, we could
just start to calculate the bound states, considering the interactions
without corrections to the Green function. However, one has to take
into account 
%---   the corrections
that the fermion-fermion interaction changes the effective mass of the
quarks and this in its turn will change the bound states considerably,
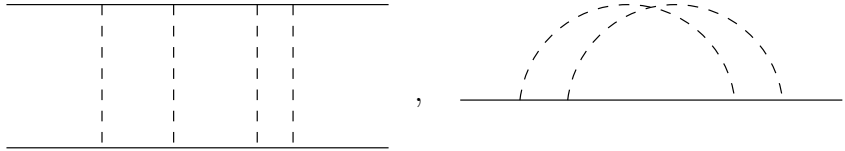
\[  \begin{picture}(0,0)%
\epsfig{file=O7.pstex}%
\end{picture}%
\setlength{\unitlength}{0.00083300in}%
\begingroup\makeatletter\ifx\SetFigFont\undefined%
\gdef\SetFigFont#1#2#3#4#5{%
  \reset@font\fontsize{#1}{#2pt}%
  \fontfamily{#3}\fontseries{#4}\fontshape{#5}%
  \selectfont}%
\fi\endgroup%
\begin{picture}(5274,924)(589,-673)
\put(3151,-361){\makebox(0,0)[lb]{\smash{\SetFigFont{12}{14.4}{\rmdefault}{\mddefault}{\updefault},}}}
\end{picture}
  \]
%---   too, 
which makes the problem more complicated. We 
thus 
will have to consider
%---   not only 
bound states and the Green functions
on equal footing. 
%---   With the change
%---   of the Green function the effective masses of the quarks will change
%---   and become large, and as a result, the bound states will change very
%---   strongly.

Up to now, the only approach which deals with this problem and is
self-consistent is the Nambu - Jona-Lasinio model [3]. It considers the
fermion Green function
%---   quark and
corrections 
%---  in the 
due to a four-fermion interaction:
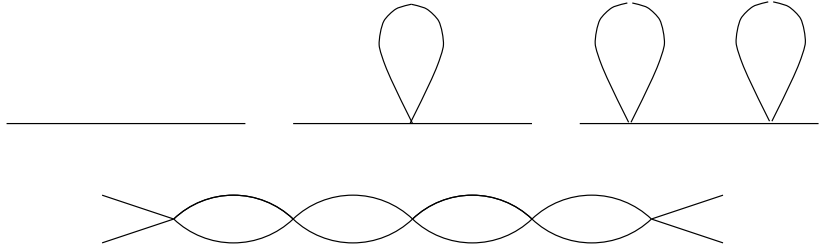
\[  \begin{picture}(0,0)%
\epsfig{file=O8.pstex}%
\end{picture}%
\setlength{\unitlength}{0.00083300in}%
\begingroup\makeatletter\ifx\SetFigFont\undefined%
\gdef\SetFigFont#1#2#3#4#5{%
  \reset@font\fontsize{#1}{#2pt}%
  \fontfamily{#3}\fontseries{#4}\fontshape{#5}%
  \selectfont}%
\fi\endgroup%
\begin{picture}(5124,1535)(589,-1123)
\end{picture}
  \]
In spite of the strong dependence on the cut-off, the model preserves
all symmetries in the Green function and in two-particle interactions.
Let us present the result of Nambu and Jona-Lasinio in a way somewhat
different from what is given in [3]. We express it as the dependence
of the renormalized mass 
$m$ on the bare mass
$m_0$. 
They found that if the
effective coupling (it depends on the definition in their case)
$\frac{g^2}{2\pi}$ is less
than unity, the curve will be just the usual one:
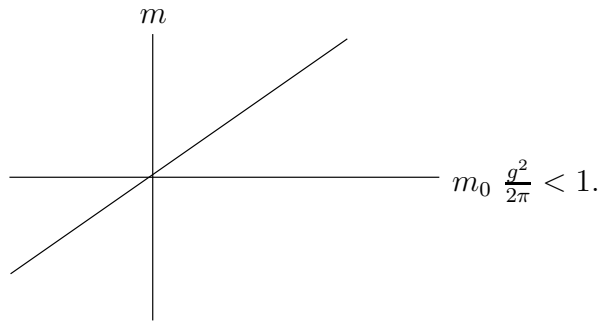
\[  \input{O9.pstex_t}  \]
%---   (here $m_0$ is the bare mass, $m$ the renormalized mass).
If, however, 
the coupling
%---   this 
$\frac{g^2}{2\pi}$ is larger than unity, the dependence
will be like 
%---   that
\footnote{This result is not always quoted, but it is
present in their paper.}:
\[  \input{O10.pstex_t}  \]
According to the interpretation of Nambu and Jona-Lasinio, the upper part
of the curve, which at $m_0=0$ reaches a finite point, corresponds to the
spontaneous symmetry breaking. But there are three solutions at $m_0=0$
and at sufficiently small $m_0$ values. What Nambu and Jona-Lasinio claim
is that the lower part of the curve is unstable, and there is a real
vacuum. I agree with this, if $m_0>m_{crit}$.
%---   ; this solution is different from what I would have here. 
We can ask: what is the source of instability
of this curve? The general argumentation is the following. Talking
about spontaneous symmetry breaking, $m$ is like a magnetic field in a
ferromagnetic; we just choose a definite direction. But $m_0$ is like an
external field, and the system is like a compass. If the external field
and the induced field are pointing at the same direction, the situation
is stable. If they point to opposite directions, the compass will change.

I am, however, not sure that the instability of the almost perturbative
solution which contains no condensate at all can be explained in such
a way. The explanation may be right for the part $b$ of the curve in
Fig.2 which corresponds to a big spontaneous magnetic moment and the
opposite direction of $m_0$. It does not work for the part $a$ close to
perturbation theory which has no spontaneous magnetic moment. And,
looking more carefully at the curve, we see that the part $b$ corresponds
to pseudoscalar states inside the Dirac see, while on the piece $a$ both
the pseudoscalar and scalar states are inside, both levels passed.
Recognizing this, one can conclude that indeed, the mentioned state is
unstable, but for a trivial reason: the corresponding level is inside
the Dirac see and it is not filled up. The problem is, what happens if
we fill up this level. It remains an open question, what can be
considered as a ground state under these conditions. And
it is a problem how to get these results in a more self-consistent
way, not depending on the cut-off so strongly.

The Nambu - Jona-Lasinio model can be reproduced in our picture. For this
purpose, just as a theoretical exercise, let us use $\alpha$  not going
to unity, i.e. draw
\[ \input{O11.pstex_t} \]
instead of the curve in Fig.1. In this case there will be a second scale
$\lambda_2$, and outside this scale we will have just point-like
interaction. This reminds of 
the Nambu - Jona-Lasinio picture, which,
apparently, can be reached somehow in our approach. The problem is, how
to write constructively the corresponding equation, and whether this
can be done at all. Of course, this constructive part can be only
approximate. But if we recognize that it can be written, then we will
be able to develop a theory in which we put the main ingredient of our
discussion as an input into our solution and try to find the real
construction. The main difference will appear in the analytic properties
of the Green function. The Green function of a fermion for such a case
would be quite different in 
its
%---   the 
analytic structure compared to the usual one.

I am afraid I will not have the time to come to this point today, but
I would like to explain just the physics.

How to write the equation? What happens in the real case and how to
deal with it? Let us start with with the Green function. What do we
know, what are we supposed to know about this Green function as a
function of $q^2$ ?
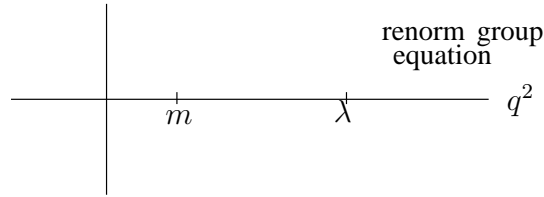
\[ \input{unnamed.pstex_t} \]
%---   If there is 
Beyond a certain
$\lambda$ in the region where the coupling is small, it
is asymptotically free; here the Green function has to satisfy the
renorm group equation. But if as a result of the interaction a mass is
acquired, this mass would be somewhere at smaller $q^2$; 
here the equation
%---   here 
becomes essentially very complicated and we are not able to extract
a reasonable structure.

The idea to write an equation which is correct in both regions, near
the threshold and at large $q^2$, and to match these two solutions,
comes from the following consideration. Suppose that $\alpha_{crit}/\pi$
is small:
\[   \frac{\alpha_{crit}}{\pi} \approx 1-\sqrt{\frac{2}{3}} = 0.2 .\]
Now, however, we may ask: how could new masses, new solutions etc.
appear at all at such a small $\alpha$. Obviously, this $\alpha$ has
to be multiplied by something large. What happens, for example, at large
$q$? We know, that there is always a logarithm of $q^2/\lambda^2$ and the
real parameter becomes
\[ \alpha_0 \ln \frac{q^2}{\lambda^2} \]
which is, in spite of the smallness of $\alpha$, big enough to change
the Green function essentially. But near the threshold there is also a
logarithm:
\[  \ln \frac{q^2-m^2}{\tilde{\lambda}^2}\,,\qquad 
        \mbox{ with some scale }\tilde{\lambda}= \lambda \mbox{ or } m  \]
which is always present. In other words, in this region there could be
also a quantity which changes seriously in spite of the relative
smallness of $\alpha$. Hence, we want to write the equation which is
correct near the threshold, taking into account correctly the singularity
of a supposed mass, and after that compare this with the renorm group
equation; we shall see whether it is possible to write an equation which
is correct in both regions, and if yes, we will try to solve it.
In order to get the singularity correctly, we take the second derivative
of $G^{-1}$ with respect to the momenta.
\begin{equation}
\label{2}
 \input{O13.pstex_t}
\end{equation}
The contribution of the first term is trivial, the second derivative of
$q+m$ gives zero. Taking the second derivative of the first graph, it can
be easily seen that
\begin{equation}
\label{3}
\partial^2 \frac{1}{(q-q')^2-i\varepsilon} = -4\pi i\delta^{(4)}(q-q')  .
\end{equation}
This gives for the first diagram $\gamma_{\mu}G(q)\gamma_{\mu}
\frac{\alpha}{\pi}$ -- just by direct calculation. In other words, we
make it local.  From this diagram we take the contribution where $k
\equiv (q-q')$ -- the momentum of the photon -- essentially equals zero.

Let us now look at the second diagram. We have here the choice of taking
the second derivative at one of the photon lines, or to differentiate
once at one line and once at the second line. Having in mind that all
the integrations would have a structure which need some logarithmic
enhancement, it would mean that the most important regions of integration
in this integral would be those where $k_1 \ll k_2$ and $k_2 \ll k_1$.
We take the derivative at $k_1$ and then set it to be zero, but for $k_2$ the
integration will give the same as before. If this integration gives us
two logarithms, we kill one and recover it after the integration of our
differential equation; but we still have the first one. But if we
differentiate once one line and once the other, we will always sit on
the region $k_1 \sim k_2$, because they have to be of the same order.
And, in this case, there is no logarithm at all; after the integration,
we will recover one, but one order will be lost. Clearly, a possible
approach is to try not to choose different diagrams, but to use the small
$k$ region of integration. Ordering the
integration inside the diagram in such a way that one momentum is much
smaller than the others, and differentiating this diagram, we will find
a relatively simple answer. Indeed, suppose that we have any diagrams
with any loops. If we differentiate some lines twice (it can be any line)
and neglect all first derivatives, we get an amplitude of the following
structure:
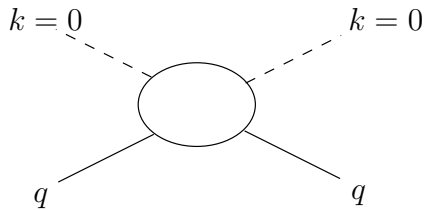
\[ \input{O14.pstex_t}  \]
This is just the Compton scattering of a zero momentum photon $k=0$, and
for this quantity the most singular contribution is obviously
\[  \Gamma_{\mu}(0,q)G(q)\Gamma_{\mu}(0,q)  \]
which corresponds to the diagram
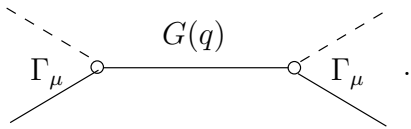
\[ \input{O15.pstex_t} .\]

%---   \[ \picbox{O14.pstex_t} = \Gamma_{\mu}(0,q)G(q)\Gamma_{\mu}(0,q) =
%---   \picbox{O15.pstex_t} .\]
But 
%---    this 
the vertex $\Gamma$ is at zero momentum 
%---   $\Gamma$, i.e.
and hence $\Gamma_{\mu}(0,q) = \partial_{\mu}G^{-1}(q)$. 
In this approximation we
can write a very simple equation:
\begin{equation}
\label{4}
\partial^2 G^{-1}(q) = \frac{\alpha(q)}{\pi}\partial_{\mu}G^{-1}(q)G(q)
 \partial_{\mu}G^{-1}(q)
\end{equation}
which differs essentially from any Bethe-Salpeter type equation. Indeed,
using a Bethe-Salpeter type equation, we do not change the vertex part
and end up with rather bad properties. Equation (\ref{4}) is scale
invariant, it is $\gamma_5$-invariant, it has many nice symmetry
properties and, what is most important, it has a correct behaviour near
the threshold.

%Of course, it would be correct to sum the infrared singularities.
The gauge is fixed, because we used
\[  D_{\mu\nu} = \frac{\delta_{\mu\nu}}{q^2}.  \]
It is an important question, what we would get in different gauges.
In Feynman gauge we are very lucky: we find an expression which does
not depend explicitly on the expression for the Green function.
Using a different gauge, we would find the infrared behaviour of this
diagram to be more complicated, and we would not be able to extract
universally the region of small $k$. We would have integrals over $q$
which are also possible to use, but with the necessity to think about the
behaviour near the threshold.

We, however, have chosen this gauge; we did not destroy the general
features and used the current conservation which just corresponds to
$\Gamma_{\mu} = \partial_{\mu}G^{-1} $. Accepting this, we can now ask,
what is the relation to the renorm group equation.

Suppose that we would like to write the renorm group equation in the
same spirit. Let us take again the second derivatives. In this case
we would be definitely correct, because we know that it is a logarithmic
approximation.

In our logarithmic approximation we will do exactly the same with the
only difference that $\alpha$ would be $\alpha(q^2)$ .
In the renorm group equation $\alpha$ is a function of $q^2$. But, of
course, $\alpha$ in general depends on two momenta: $k^2$ and $q^2$.
And in the renorm group equation at large 
%---   distances
momenta, in the ultraviolet
region, $\alpha$ depends on the variable which is the largest.
Since we are close to $k=0$, this means that here we will have
$\alpha(q^2)$, 
and we will recover the renorm
group equation at large $q^2$. If we solve this equation with a
slowly varying $\alpha$, we will be correct in both the threshold region
and the ultraviolet region.

%---   To see the connection between different 
%---   solutions for the Green function,
We also have to formulate an equation for the bound states under the same
assumption. Looking for bound states, we consider scalar and pseudoscalar
vertices. This vertex
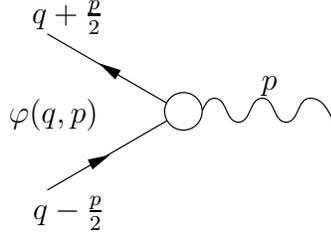
\[ \input{O16.pstex_t}   \]
has to be equal to 
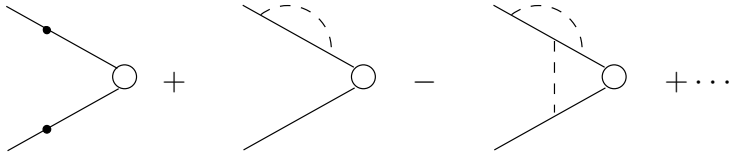
\[ \input{O17.pstex_t}   \]

Here $\varphi(q,p)$ depends on $p$, the total momentum of a pair, and
  $q$, the quark momenta being given by $q+p/2$ and
$q-p/2$. 
%Using the same trick as before, and taking the second derivative
%of $\varphi$ as a function of $q$ and $p$, what will we find?
%If we take the second derivative, our line becomes a $\delta$-function.
%And the contribution from this diagram will be the following: it is easy
%to show that now it contains minus, $\Gamma_{\mu}$, $G$, the same
%$\varphi$, $G$, $\Gamma_{\mu}$.
%The product $G_{\mu}\partial_{\mu}G^{-1}G$ reminds $A_{\mu}\varphi
%A_{\mu}$ in the sense that the position of $G$ is preserved.
%But this is not the only contribution.
%And it is easy to show that the contribution is the following:
With the same procedure as in obtaining the equation for the Green
function we find for the vertex (see [4] for some details):
\begin{equation}
\label{5}
\partial^2 \varphi(q,p) = \frac{\alpha}{\pi}\left[ A_{\mu}(q)\partial_{\mu}
 \varphi(q,p) + \partial_{\mu}\varphi\tilde{A_{\mu}}(q) - A_{\mu}(q)\varphi(q,p) \tilde{A_{\mu}}(q)\right]\,,
\end{equation}
where $A_{\mu}=\partial_{\mu}G^{-1}G$, $\tilde A_{\mu}= G
\partial_{\mu}G^{-1}$\,.
It means that we have two equations in this approximation. We used this
approximation just to be constructive and to study what will result if we
make this approximation. In principle, solving both equations 
we will get everything
what is necessary: we know $G$ and $A_{\mu}$, we have a linear equation
for bound states, we can see what is the type of the energy etc. The
equation for the bound states has very nice features.  It is beautiful
from the point of view of the Goldstone theorem in the following
sense.

Suppose I have some symmetry in my equation,
e.g. $\gamma_5$-invariance. Since there is no $\gamma_5$ in
equation (5), it is  $\gamma_5$-invariant. But of course the boundary
condition for $G^{-1}$ at $q \to \infty$ is just $G_0^{-1}=(\hat q -
m_0)$, and thus destroys the symmetry. But suppose that $m_0=0$. In
this case there would be symmetry here, which means that the Green
function will not be unique, since it can be 
\begin{displaymath}
  G^{-1} + \delta G^{-1}, 
\end{displaymath}
where $G^{-1}$ is some solution and $ \delta G^{-1} \propto \gamma_5
G^{-1}$\,. This means that the variation  $\delta G^{-1}$ also is
important. What would be the equation for the variation? If we
calculate the variation on both sides of equation (4) we obtain
\begin{displaymath}
  \partial_\mu\,(\delta G^{-1}) = \frac{\alpha}{\pi}\left(
    \partial_\mu(\delta G^{-1})G\partial_\mu G^{-1} +
        \partial_\mu G^{-1} G \partial_\mu (\delta G^{-1}) -
        \partial_\mu G^{-1} G (\delta G^{-1}) G \partial_\mu G^{-1}
    \right)   \,,
\end{displaymath}
so we find that $\varphi=\delta G^{-1}$ fulfils equation (5) at
$p=0$. It means that if some symmetry is broken, i.e. if there are
multiple solutions of the equation for the Green function, we always
will have some solution of the equation for the vertex at $p=0$,
which is the Goldstone.

It is clear, that in the present model we can discuss many questions,
use a running coupling $\alpha$ as in Fig. 1 and reproduce the
NJL-features without any essentials depending on a cutoff.
Before discussing this point further, we will first look for the
solution of (4) and discuss the result. 

Above, we introduced $A_\mu = (\partial_\mu G^{-1}) G$, which is a very
useful quantity. Since $G^{-1}=a\frac{\hat q}{q}+b$ is essentially a
$2 \times 2$ matrix, $A_\mu$ is a $U(2)$ gauge potential: 
\begin{equation}
\label{6}
  \partial^2G^{-1} = \partial_\mu((\partial_\mu G^{-1})\,G\,G^{-1}) 
  = (\partial_\mu A_\mu)G^{-1} + A_\mu (\partial_\mu G^{-1}) =
  \frac{\alpha}{\pi}A_\mu\partial_\mu G^{-1}
\end{equation}
where in the last step we used Eq. (4). Multiplying from the right by
$G$ we thus find
\begin{equation}
\label{7}
\partial_{\mu}A_{\mu} + A_{\mu}A_{\mu} = \frac{\alpha}{\pi}A_{\mu}A_{\mu}.
\end{equation}
This means that
\[\partial_{\mu}A_{\mu}=-\beta A_{\mu}A_{\mu};\]
and $A_\mu$ is a pure $U(2)$-gauge potential with a condition $\beta = 1-\frac{\alpha}{\pi}$.
Of course, this is just a useful trick. Important is to write down the real
equation for the Green function. The most natural thing is to express
$G^{-1}$ in the form
\[   G^{-1} = \rho\, e^{\frac{\varphi}{2}\hat{n}} ,\]
where $\hat{n}$ is a $2 \times 2$-matrix
\[  \hat{n} = \frac{\hat{q}}{q} .\]
It is just easier to use this form for our purpose: we can find an
equation for $\rho$ and an equation for $\varphi$. Both are functions of
$q^2$: $\rho(q^2)$, $\varphi(q^2)$. There is, however, no scale in the
equation; it contains only a derivative of $q$. If we introduce
\[  \xi = \ln q ,\]
we will find an equation in which $\xi$ can be considered as ''time'', and
which is an oscillator equation. In fact there are two oscillators, one
for $\rho$, the other for $\varphi$, and they will satisfy non-linear
equations. For $\varphi$ we find
\begin{equation}
\label{8}
\ddot{\varphi} + 2\left(1+\beta\frac{\dot{\rho}}{\rho}\right)\dot{\varphi}
 - 3\sinh\varphi = 0 .
\end{equation}
This is just an oscillator with damping; a similar equation can be
written for $\rho$. Important is that that there has to be ''energy''
conservation in this equation. Indeed, we said that $\xi$ plays the
role of time; it, however, did not enter the equation explicitly.
Thus there has to be a conservation law which, as it is easy to show,
leads to
\begin{equation}
\label{9}
\left(1+\beta \frac{\dot{\rho}}{\rho}\right)^2 =  1 + 
 \beta^2\left(\frac{\dot{\varphi}^2}{4}-3\sinh^2 \frac{\varphi}{2}\right) .
\end{equation}
We thus can eliminate $\rho$ altogether, and find the equation for
$\varphi$
\begin{displaymath}
\ddot{\varphi} +
 2\sqrt{1-\beta^2\left(3\sinh^2\!\frac{\varphi}{2}-\frac{\dot\varphi^2}{4}\right)}\dot{\varphi}
 - 3\sinh\varphi = 0\,,
\end{displaymath}
which is an oscillator with damping. Having this in mind is sufficient
to understand the structure of the solution. Indeed, what is this
$\varphi$ ? We have 
\begin{equation}
\label{10}
G^{-1} = \rho\cosh\frac{\varphi}{2} + \frac{\hat{q}}{q}\rho\sinh
 \frac{\varphi}{2} .
\end{equation}
The perturbative solution is $\varphi$ close to $i\pi$. In this case
the first term is zero, the other is proportional to $\hat{q}/q$ -
this corresponds to the massless solution. Since $m_0$ is small, we
have to have solutions like this at $q \rightarrow \infty$.

Now we have to find the solution everywhere. Let us first investigate
the equation without damping; we get
\[ \ddot{\varphi} - 3\sinh\varphi = 0 .\]
If we go to the Euclidean space, $\varphi = i\psi$, the potential becomes
a periodical potential:
\[ \ddot{\psi} - 3\sin\psi = 0 .\]
\[ \input{O18.pstex_t}    \]
We have to look for a possible solution for this
structure with damping. What does this mean? For the oscillator
with damping any solution at $\xi \rightarrow \infty$ has to be in a
minimum, because the energy is decreasing. But if $\xi$ is going
to $-\infty$, the energy is growing. What could be in this case a normal,
reasonable solution? It is almost clear that the only possibility is
to put at $\xi \rightarrow -\infty$ the ''particles'' in this
oscillator at the maximum,
and start to move them slowly; eventually, they will appear inside the
well.

%\begin{equation}
%\label{11}
%\left(1+\beta\frac{\dot{\rho}}{\rho}\right)^2 = \beta^2 \left[\frac
% {\dot{\varphi}^2}{4} + 1 - 3\sinh^2 \frac{\varphi}{2} \right] = 4\beta^2
%\end{equation}
%We recover the value of $\alpha$:

%AGAIN, FROM HERE ON I HAD NO TEXT

%\[ \nu = \sqrt{1+3\beta} \pm \sqrt{3\beta^2-2} \]
%\[ \sqrt{\frac{2}{3}} = \beta = 1 - \frac{\alpha}{\pi}  \]
%\[ \frac{\alpha}{\pi} = 1 - \sqrt{\frac{2}{3}} \]

There is a most important question, namely: what is the critical
coupling in this case? What do we know about an oscillator with
damping? If the damping is large enough, all the trajectories will go
monotonically to the minimum. If the damping is sufficiently small,
the solution will start to oscillate. In order to see when this
transition happens, we have to look for the equation just near the
minimum $\psi=\pi$. With $\phi \equiv \pi - \psi$ we have for small
$\phi$
\begin{displaymath}
  \ddot \phi + 2 \sqrt{1+3\beta^2}\dot\phi + 3 \phi = 0\,,
\end{displaymath}
with fundamental solutions
\begin{displaymath}
  \phi_{1,2} = \mbox{e}^{\nu_{1,2}\xi} \mbox{ where } \nu_{1,2} =
  -\sqrt{1+3\beta^2}\pm\sqrt{3\beta^2-2}\,.
\end{displaymath}
So we have monotonic behaviour for $3\beta^2-2>0$. On the other hand
if $\beta^2<\frac{2}{3}$, i.e.
\begin{displaymath}
  \frac{\alpha_{crit}}{\pi} = 1 - \sqrt{\frac{2}{3}} <
  \frac{\alpha}{\pi} <  1 + \sqrt{\frac{2}{3}}\,,
\end{displaymath}
we will have oscillations before reaching the minimum. The critical angle $\psi_c$, which
separates the regions where the solution is monotonic and where it
oscillates can be
shown (see e.g. (4)) to be given by 
\begin{displaymath}
  \sin^2 \frac{\psi}{2} = \left(\frac{2}{3}-\beta^2\right)
  \sqrt{\frac{1+3\beta^2}{1-\beta^2}}
  \frac{1}{1 + \sqrt{(1+3\beta^2)(1-\beta^2)}}\,.
\end{displaymath}
\[  \input{O19.pstex_t}  \]
%\[  \picbox{O20.pstex_t}  \]

Up to now we have considered a constant coupling $\alpha$. We know that
for $q>\lambda$ the Green function is determined by perturbation
theory, which has to match the solutions in the region of smaller
$q$. If $\beta^2>\frac{2}{3}$ for all $q^2$, the solution which goes
as $\psi \approx \frac{q}{m_c}$ for $q\to 0$ matches the solution
$\frac{i}{2}(\psi-\pi) \approx \frac{m_0}{q} + \frac{\nu_1^2}{q^3}$ for
$q\to \infty$ monotonically. This determines $m_0$ as a function of $m_c$ in a
unique way.
Let $\lambda$ be the value of $q$ where
$\beta^2(\lambda^2)=\frac{2}{3}$.
If, however, $\beta^2<\frac{2}{3}$ below $q=\lambda$, the solutions can
oscillate and $m_0(m_{c_i})=0$ for some $m_{c_i}$ as indicated in
Fig.6. 
%--- Matching with perturbation theory:
%--- The result is
\[  \begin{picture}(0,0)%
\epsfig{file=O21.pstex}%
\end{picture}%
\setlength{\unitlength}{0.00083300in}%
\begingroup\makeatletter\ifx\SetFigFont\undefined%
\gdef\SetFigFont#1#2#3#4#5{%
  \reset@font\fontsize{#1}{#2pt}%
  \fontfamily{#3}\fontseries{#4}\fontshape{#5}%
  \selectfont}%
\fi\endgroup%
\begin{picture}(2479,3219)(1489,-2473)
\put(2626,614){\makebox(0,0)[lb]{\smash{\SetFigFont{12}{14.4}{\rmdefault}{\mddefault}{\updefault}$m$}}}
\put(3968,-1043){\makebox(0,0)[lb]{\smash{\SetFigFont{12}{14.4}{\rmdefault}{\mddefault}{\updefault}$m_0$}}}
\end{picture}
  \]
  \[   Fig.6 \]
 This then is a
solution corresponding to broken chiral symmetry.

\section*{References}
\begin{description}
\item[1] V. N. Gribov, Orsay lectures on confinement (I), 
preprint  LPTHE
 Orsay 92-60 (1993); hep-ph/9403218
\item[2] V. N. Gribov, Orsay lectures on confinement (II), preprint LPTHE
 Orsay 94-20 (1994); hep-ph/9407269
\item[3] Y. Nambu, G. Jona-Lasinio, Phys. Rev. \underline{122} (1965), 345
\item[4] V. N. Gribov, Lund preprint LU 91-7 (1991)
\item[5] V. N. Gribov, QCD at large and short distances, 
  Bonn preprint TK 97-08 (1997), hep-ph/9807224.
\item[6] V. N. Gribov, The theory of quark confinement, Bonn preprint
 TK 98-09 (1998), hep-ph/9902279
\end{description}
\end{document}